\documentclass[
    ,final            
  ]
  {aipproc}

\layoutstyle{8x11single}

\newcommand{\gsim}{\lower .75ex \hbox{$\sim$} \llap{\raise .27ex \hbox{$>$}} }
\newcommand{\lsim}{\lower .75ex \hbox{$\sim$} \llap{\raise .27ex \hbox{$<$}} }

\begin{document}

\title{Instabilities in dark coupled models and\\
  constraints from cosmological data}


\author{Laura Lopez Honorez}{
  address={Departamento de F\'\i sica Te\'orica \& Instituto de F\'\i sica Te\'orica,\\
Universidad Aut\'onoma de Madrid, 28049 Cantoblanco, Madrid, Spain and\\
 Service de Physique Th\'eorique,
Universit\'e Libre de Bruxelles, 1050 Brussels, Belgium}
}
\author{ Olga Mena}{
  address={Instituto de F\'{\i}sica Corpuscular, IFIC, CSIC and
Universidad de Valencia, Spain}
}

\begin{abstract}
  Coupled dark matter-dark energy systems can suffer from non-adiabatic instabilities at early
  times and  large scales. In these
  proceedings, we consider two
  parameterizations of the dark sector interaction. In the first one
  the energy-momentum transfer 4-vector is parallel to the dark
  matter 4-velocity and in the second one  to the dark
  energy 4-velocity. In these cases, coupled models which suffer from
  non-adiabatic instabilities can be identified as a function of a
  generic coupling $Q$ and of the dark energy equation of state
  $w$. In our analysis,  we do not refer to any particular cosmic
  field.  
  We  confront then a viable class of models in which the
  interaction  is directly proportional to the dark energy density  and to the
  Hubble rate parameter to recent cosmological data. In that framework, we show that
  correlations between  the dark coupling and several cosmological
  parameters allow for a larger neutrino mass than
  in uncoupled models. 
  
\end{abstract}

\maketitle

\section{Introduction}
\label{subsec:intro}

Interactions between dark matter and dark energy  are still allowed by
observational data today. At the level of the background evolution
equations, one can generally introduce a coupling
between these two sectors   as follows:
\begin{eqnarray}
  \label{eq:EOMm}
  \dot\rho_{dm}+ 3\mathcal{H}\rho_{dm} &=& Q\,,\\
\label{eq:EOMe}
 \dot\rho_{de}+ 3 \mathcal{H}\rho_{de}(1+ w)&=&- Q\,.
\end{eqnarray}
$\rho_{dm} (\rho_{de})$ denotes the dark matter (dark energy) energy density,  the dot indicates derivative with
respect to conformal time $d\tau = dt/a$, $\mathcal{H}= {\dot a}/a$
and  $w=P_{de}/\rho_{de}$ is the dark-energy equation of state ($P$
denotes the pressure).
We work with the Friedman-Robertson-Walker (FRW) metric,
assuming a flat universe and  pressureless dark matter $w_{dm} =
P_{dm}/\rho_{dm}=0$.

\begin{figure}[t]
\includegraphics[width=6cm]{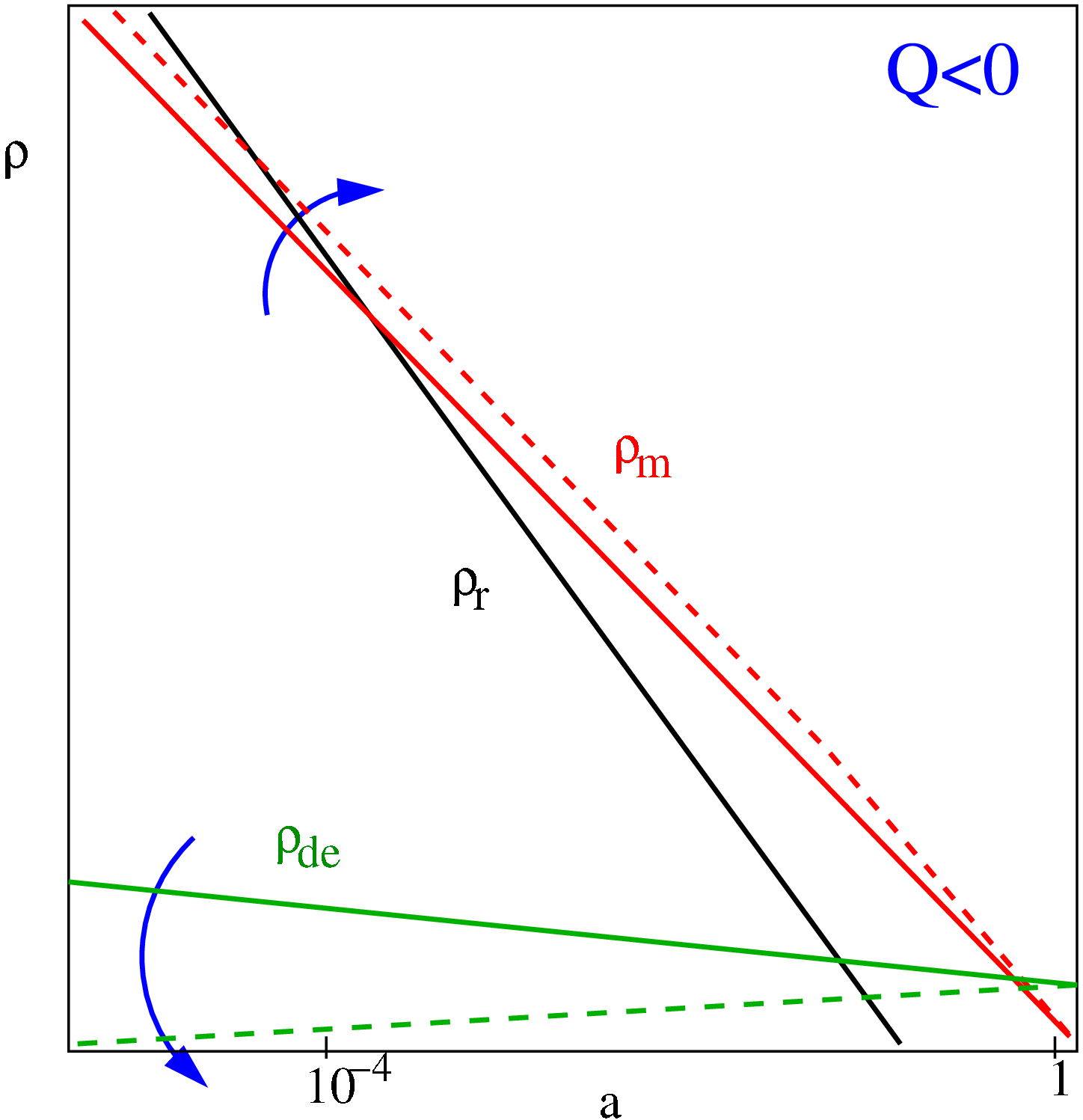} 
\caption{\it This figure illustrates roughly the effect of a negative dark coupling term $Q$ on dark
  matter and dark energy evolution. $a$ is the scale factor,
  $\rho_i$ with $i=m,r,de $ corresponds  to matter (dark matter plus
  baryon), radiation  and dark energy densities respectively. Solid curve
  account for an uncoupled model with constant dark energy equation of state
  $w>-1$ and dotted curve for coupled models with $Q<0$ with $w>-1$.      }
\label{fig:Q}
\end{figure}

Q encodes the dark coupling and drives the energy exchange between dark
matter and dark energy. For {\it e.g.} $Q<0$ the energy flows from
dark matter to dark energy. It also changes the  dark
matter and dark energy redshift dependence acting as an extra
contribution to their effective equation of state. For
{\it e.g.} $Q<0$, dark matter redshifts faster, as a consequence, there is  more
dark matter in the past compared to uncoupled
scenarios  assuming that the dark matter
density today is the same in the two models (see also the discussion in
Ref.~\cite{CalderaCabral:2009ja}). This general feature of coupled
models is sketched in Fig.~\ref{fig:Q} (for a particular form of $Q$,
see also Fig.~\ref{fig:fishesplots}).


 Also notice that a universe in accelerated expansion today requires $w<-1/3$ even in the presence 
of a dark coupling. Indeed, the deceleration parameter satisfies
\begin{equation}
  \label{eq:decel}
q=-  \frac{\dot {\mathcal H} }{{\mathcal H}^2}=\frac12 (1+3\,w\,\Omega_{de})~,
\end{equation}
either with or without dark coupling. The curvature contribution has
been neglected in this equation. 

In order to deduce the evolution of  density and velocity perturbations in  coupled models, we need an
expression of the energy transfer in terms of the stress-energy tensor: 
\begin{eqnarray}
\nabla_\mu T^\mu_{(dm)\nu} =Q_\nu \quad\mbox{and}\quad
\nabla_\mu T^\mu_{(de)\nu} =-Q_\nu.
\label{eq:conservDMDE}
\end{eqnarray}
The  4-vector $Q_\nu$ governs the  energy-momentum transfer between the dark
components and  $T^\mu_{(dm)\nu}$ and $T^\mu_{(de)\nu}$ are the energy-momentum tensors for the dark matter 
and dark energy, respectively. Equation~(\ref{eq:conservDMDE}) guaranties the
conservation of the total energy-momentum tensor. In the following we consider
two scenarios which reproduce at the background level Eqs.~(\ref{eq:EOMm}) and~(\ref{eq:EOMe}):
\begin{enumerate}
\item $Q_\nu$ is parallel to the dark matter
  four velocity $u_{\nu}^{(dm)}$:
  \begin{equation}
    Q_\nu = Q u_{\nu}^{(dm)}/a.
\label{eq:um}
  \end{equation}
This choice of parameterization, which was first proposed in
Ref.~\cite{Valiviita:2008iv},  avoids momentum transfer in the rest frame of dark 
matter 
({\it i.e} $v^i_{(dm)}=0$), 
\item $Q_\nu$ is parallel to the dark energy
  four velocity $u_{\nu}^{(de)}$:
  \begin{equation}
    Q_\nu = Q u_{\nu}^{(de)}/a.
\label{eq:ue}
  \end{equation}   
In this case the dark matter velocity (Euler)
  equation is modified by the coupling and the weak equivalence principle is
  violated (see {\it e.g.} Ref~\cite{Koyama:2009gd} for a recent
  discussion). Notice that one example for this interaction  is 
$Q_\nu= \beta \rho_{dm} \nabla_\nu \phi$ \cite{Wetterich:1994bg,Amendola:1999qq}, where 
 $\phi$ would be  a coupled  dark energy scalar field and 
 $u_{\nu}^{(de)}\propto \nabla_\nu \phi$.   
\end{enumerate}

It was first pointed out in Ref.~\cite{Valiviita:2008iv} 
that the dark coupling terms which appears in the non-adiabatic dark
energy pressure perturbations are a source for  early time
 instabilities  at large scales for coupled models satisfying
 Eqs.~(\ref{eq:conservDMDE}) and~(\ref{eq:um}). 
Their analysis was however restricted to $Q>0$ proportional to the
 dark matter density, in which case the coupled model is particularly unstable
 for a constant dark energy equation of state.
Non-adiabatic instabilities were subsequently  analyzed  by several authors, 
and it was proved that the stability depends on the type of dark coupling $Q$, on the dark energy equation of state
$w$ and on the $Q_\nu$ 4-velocity dependence,   see
Ref.~\cite{Jackson:2009mz,He:2008si,Gavela:2009cy,Majerotto:2009np} (for similar instabilities pointed out in
coupled quintessence models, see Ref.~\cite{Corasaniti:2008kx,Chongchitnan:2008ry}).

Among these references, we proposed in Ref.~\cite{Gavela:2009cy}, a criteria associated 
to the dubbed {\bf \it doom factor}  to identify the stability region
of coupled models with a constant dark energy equation of state $w$
satisfying Eqs.~(\ref{eq:conservDMDE}) and~(\ref{eq:um}).
The doom factor is a function of the model parameters such as $Q$ and $w$, but it is
defined independently of the explicit form of the coupling $Q$. We review
this result in the next section and extend it to  models satisfying Eqs.~(\ref{eq:conservDMDE}) and~(\ref{eq:ue}). 

Based on this analysis, we study the compatibility  of  a successful class of models, in which $Q$ is proportional 
to the dark energy density, with a fixed dataset including WMAP 5 year~\cite{Dunkley:2008ie,Komatsu:2008hk} , HST~\cite{Freedman:2000cf}, SN~\cite{Kowalski:2008ez}, $H(z)$~\cite{Simon:2004tf} and LSS~\cite{Tegmark:2006az} data.
This work was done using the publicly available \texttt{CAMB}
code~\cite{Lewis:1999bs} and \texttt{cosmomc} package~\cite{Lewis:2002ah}.
The latter were  modified in order to include the interaction between the dark
matter and dark energy components and the modified dark matter velocity
equation  in the case of  coupled models satisfying Eq.~(\ref{eq:ue}) (see Eq.~(\ref{eq:thetaees})).
Present data will be shown to allow for a sizeable interaction strength and to
imply weaker cosmological limits  on 
neutrino masses with respect to non-interacting
scenarios. 
\section{Origin of non-adiabatic instabilities}
\label{sec:instab}

Non-adiabatic instabilities arise at linear order in perturbations and appear
to be driven by the dark coupling term present in the non-adiabatic dark
energy pressure perturbation~\cite{Valiviita:2008iv}. Using the definitions
of gauge invariant density perturbation and entropy perturbation (see {\it
  e.g.} Ref.~\cite{Kodama:1985bj}), one can work
out a general expression which relates the dark energy pressure perturbation $\delta
P_{de}$ in its rest frame to the one in any other frame. The former is 
characterized by   $\hat c_{s\,de}^2 =[\delta P_{de}/\delta \rho_{de}]|_{\rm rf}$, the propagation speed of pressure fluctuations in the
rest frame of dark energy which has to be distinguished from   $c_{a\,de}^2=
\dot P_{de}/\dot \rho_{de}$, the so called ``adiabatic sound speed''. In the
synchronous or the Newtonian gauge, one obtains:                                   \begin{eqnarray}
{\delta P_{de}}&=&\hat c_{s\,de}^2 \delta \rho_{de} +(\hat c_{s\,de}^2-
c_{a\,de}^2) \dot \rho_{de} \frac{\theta_{de}}{k^2}\,,
 \label{eq:dpcs}
\end{eqnarray}
where $\delta \rho_{de}$ denotes the dark  energy density
perturbation and   $\theta_{de} \equiv
\partial_i v^i_{(de)}$ is the divergence of the dark energy proper velocity,
$v^i_{(de)}$.
Using equation (\ref{eq:EOMe}), we see that the dark coupling
resulting from the $\dot \rho_{de}$ term 
directly affects the ${\delta P_{de}}$. In Eq.~(\ref{eq:dpcs2}), we rewrite
Eq.~(\ref{eq:dpcs}) in term of the dubbed doom factor ${\bf
  d}\propto Q$ which is a useful tool to spot the combined role of $Q$ and $w$
in driving the non-adiabatic instabilities. In the following we illustrate
this feature in the framework of coupled models satisfying Eqs.~(\ref{eq:um}) and~(\ref{eq:ue}).

Notice that non-adiabatic instabilities differ from the adiabatic ones. 
The latter appear  at relatively small scales and late
times.  In the adiabatic regime,  the effective sound speed of the fluid
 tends towards the adiabatic one which turns out to be
 negative~\cite{Afshordi:2005ym,Kaplinghat:2006jk,Bean:2007ny}. As   a 
consequence, pressure no longer counteracts the effect of gravity and
instabilities can breakout.  

In the following, we work with constant equation of state $w$ in which case
($c_{a\,de}^2=w$) and we assume that our universe is in accelerating expansion today, which
implies that $w<-1/3$. Moreover, we restrict our analysis to the case  $\hat
c_{s\,de}^2>0$ and $\hat
c_{s\,de}^2=1$ will be assumed for numerical computation.

\section{Growth equation  at large scales and the doom factor}
\label{sec:growth}
A cartoon equation of the growth equation governing the evolution of  energy density linear
perturbation for any species $i, j$ is given by:  
\begin{center}
\vspace*{-2cm}
\rotatebox{270}{\includegraphics[width=0.40\textwidth]{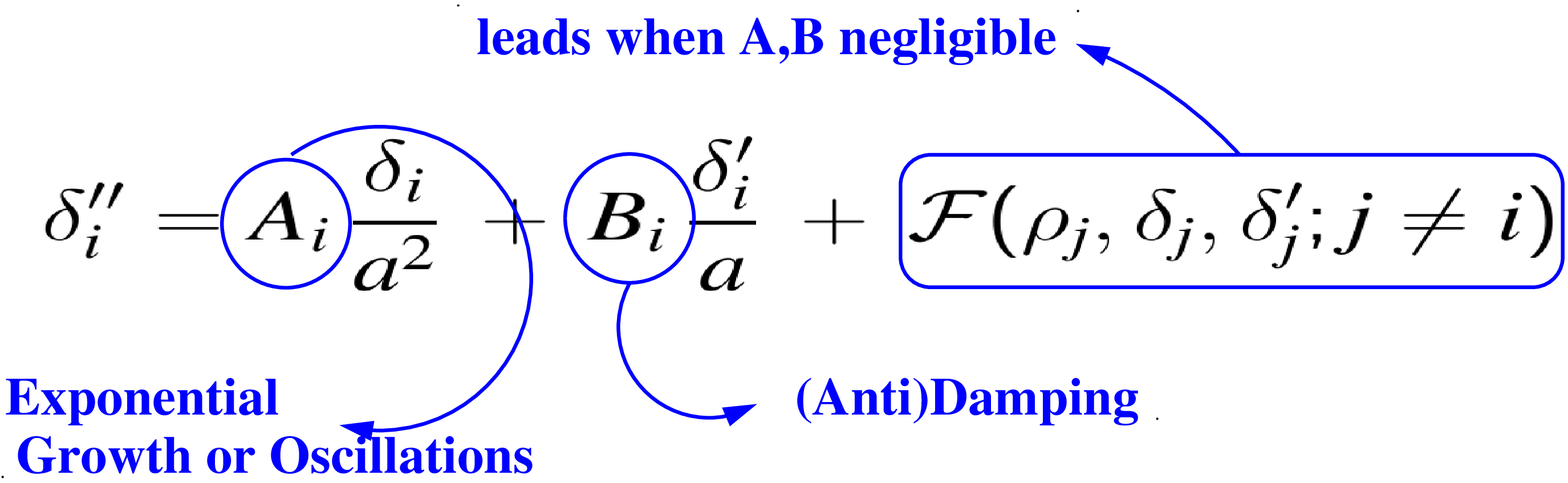}}
\end{center}
\vspace*{-1.5cm}
where $\delta_i= \delta \rho_i/\rho_i$ and the prime denotes a
derivative with respect to the scale factor $'\,= \partial/\partial \,a$.
The evolution of a perturbation depends on the relative weight of the
three terms present in this equation {\it and} on their signs: 
\begin{enumerate}
\item For positive $A$, the $A$ and $B$ terms taken by themselves would induce a rapid growth of 
the perturbation, which may be damped or antidamped (reinforced) depending on whether $B$ is 
negative or positive, respectively. In particular, for $A$ and $B$ both positive, the solution 
may enter in an exponentially growing, unstable, regime. \label{un}
\item For negative $A$, in contrast, the $A$ and $B$ terms taken alone describe a harmonic oscillator, 
with oscillations damped (antidamped) if $B$ is negative (positive). In the $A,B<0$ regime, the third 
term may play in fact the leading role. \label{dos}
  \end{enumerate}
In the standard  uncoupled scenario, the  dark matter perturbations behave as in case \ref{un} above (with $A>0$ and $B<0$), 
while the dark energy ones provide an example of behavior as in case \ref{dos}.

For coupled models, we concentrate on the case in which the dark-coupling terms
dominate over the usual one, in order to put forward the presence of non
adiabatic instabilities (see Ref.\cite{Gavela:2009cy} for more details). 
For this purpose, let us rewrite Eq.~(\ref{eq:dpcs}) as:
\begin{eqnarray}
\frac{\delta P_{de}}{\delta \rho_{de}}
 &=&\hat c_{s\,de}^2 +3(\hat c_{s\,de}^2- c_{a\,de}^2) 
 (1+w)\left(1 + {\bf d}  \right) \frac{{\mathcal H}\theta_{de}}{k^2\delta_{de}}\,,
 \label{eq:dpcs2}
\end{eqnarray}
where  ${\bf d}$ refers to the doom factor which we have defined as:
\begin{equation}
\label{eq:maldito}
 {\bf d} \equiv \frac{Q}{3\mathcal H\rho_{de}(1+w)}\,.
\end{equation}
The
{\it strong coupling regime} can be characterized by $|{\bf d}|>1$,
which also guarantees that the interaction among the two dark sectors drives the non-adiabatic 
contribution to the dark energy pressure perturbation.  

In Ref.~\cite{Gavela:2009cy}, the strong coupling regime analysis was restricted to  coupled models satisfying Eqs.~(\ref{eq:conservDMDE}) and~(\ref{eq:um}). It is however rather straightforward to
generalize this result to models satisfying Eq.~(\ref{eq:ue}) (see also the
appendix).  At large scales-early times (${\cal H}/k\gg 1$),  the leading
contributions in $Q$, or equivalently in ${\bf d}$, to the second order differential
equation for $\delta_{de}$ reads:
 \begin{eqnarray}
  \label{eq:grstrongfullours}
\delta_{de}''&\simeq&  \,3\,{\bf d}\,(\hat c_{s\,de}^2+b)\left(\,\frac{\delta_{de}'}{a} 
    \,+\,3b\frac{\delta_{de}}{a^2}\frac{(\hat c_{s\,de}^2-w)}{\hat c_{s\,de}^2+b}
     \,+\,\frac {3(1+w)}{a^2}\delta[\,{\bf d}\,]\,\right)+...
  \end{eqnarray}
where we use the $b$ notation introduced in  Ref.~\cite{Jackson:2009mz} , $b=1$ stands
for models with $Q_\nu\propto u_\nu^{(dm)}$ (\ref{eq:um}) and
$b=0$  for models with $Q_\nu\propto u_\nu^{(de)}$ (\ref{eq:ue}).
The sign of the  coefficient $B_e$ of $\delta_{de}^\prime$  in
this expression is crucial for the analysis of instabilities.  Assuming $\hat
c_{s de}^2>0$,  it reduces to the sign of the doom 
factor ${\bf d}$. As previously argued, a positive  ${\bf d}$  can  trigger large scale
instabilities.
Similar second order differential equations for $\delta_{de}$ were obtained in
Refs.~\cite{Jackson:2009mz,He:2008si} for particular expressions of the dark
coupling $Q$ and an analytical form of their solutions were
derived in order to determine when $\delta_{de}$ blows up. In particular, the
results of Ref.~\cite{He:2008si} 
confirm those of Ref.~\cite{Valiviita:2008iv}  for  positive $Q\propto
\rho_{dm}$ and $1+w>0$. In addition, our Eq.~(\ref{eq:grstrongfullours})
reproduces as well the dark coupling leading contributions of
Ref.~\cite{Jackson:2009mz}, which studied  $Q\propto\rho_{de}$ coupled models,
in particular, they carried out the first stability analysis in the framework of
$Q_\nu\propto\rho_{de} u_\nu^{(de)}$ . 
\section{Viable models: $Q\propto\rho_{de}$ }
\label{sec:rhoe}

Using the tools that we have  developed in the previous section, we can now easily verify that for
\begin{equation}
  Q=\xi {\mathcal H} \rho_{de},
\label{eq:Qrhoe}
\end{equation}
we have  rather  simple and  viable models for specific
combination of $1+w$ and of the dimensionless constant
coupling $\xi$. Indeed in these models the doom factor of Eq.~(\ref{eq:maldito}) is  given  by:
 \begin{equation}
\label{eq:maldito_us}
{\bf d }= \frac{\xi}{3(1+w)}\,.
\end{equation}
 When  ${\bf d } <0$, that
is, for $\xi<0$ and $1+w>0$ (or $\xi>0$ and $1+w<0$), no instabilities
are expected and we can safely fit the coupled models to cosmological data. 
 Notice that this result is in agreement with those  
of Refs.~\cite{He:2008si,Jackson:2009mz} which restricted though   their
stability analysis to the $\xi>0$ case.

For the sake of completeness we first give  the expressions of the
background dark fluids energy densities and first order perturbation
evolution equations which were introduced in the \texttt{CAMB}
code~\cite{Lewis:1999bs}. As previously mentioned, we work in the
synchronous gauge.

\subsection{Background}

\begin{figure}[h!]
\includegraphics[height=.4\textheight]{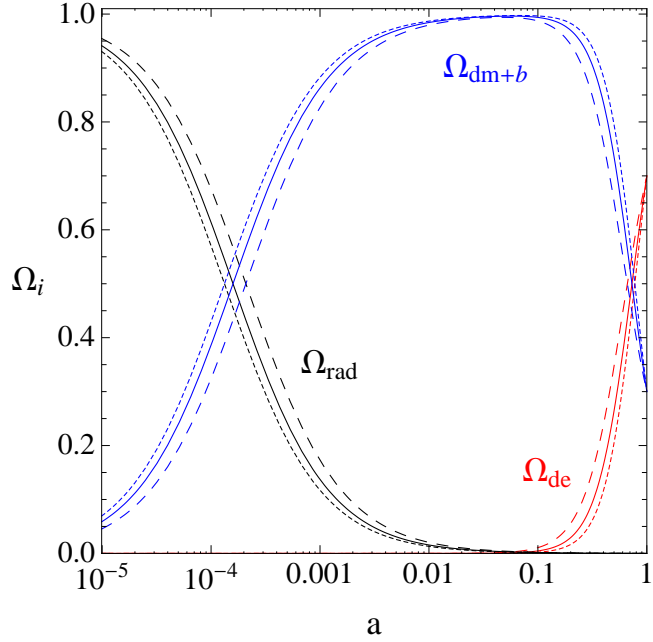}
\caption{\emph{Scenario with $Q \propto \rho_{de}$.
Relative energy densities of dark matter plus baryons $\Omega_{dm+b}$ (blue), radiation $\Omega_{rad}$ (black) 
and dark energy $\Omega_{de}$ (red), as a function of the scale factor $a$, for w=-0.9. 
Three values of the coupling are illustrated: $\xi= 0$ (solid curve), $0.25$ (long dashed curve) and
$-0.25$ (short dashed curve). }}
  \label{fig:fishesplots}
\end{figure}

The solutions to Eqs.~(\ref{eq:EOMm}) and (\ref{eq:EOMe})
using Eq~(\ref{eq:Qrhoe})  are: 
\begin{eqnarray}
 \rho_{dm}&=& \rho_{dm}^{(0)} a^{-3} + 
      \rho_{de}^{(0)}\frac{\xi}{3w+\xi}(1- a^{-3w-\xi})  a^{-3} \label{eq:adm}\,,\\
  \rho_{de}&=& \rho_{de}^{(0)}a^{-3(1+w)+\xi}\label{eq:ade}\,.
\end{eqnarray}
The dark energy density is thus always positive, all along the cosmic evolution and since its 
initial moment. To ensure that the same happens with the dark matter density, all values of 
$w<0$ are acceptable for $\xi<0$, while for positive $\xi$ it is required that 
$$\xi\lsim -w.$$ 
In Fig.~\ref{fig:fishesplots}, we see that negative (positive) couplings lead to more (less) dark matter in the past than in
the uncoupled case. In the following we focus on negative couplings in
order to avoid non-adiabatic instabilities.
%
\subsection{Linear perturbation theory}
\label{lin}

In the synchronous comoving gauge, metric scalar perturbations are described by the two usual 
fields~\cite{Ma:1995ey} $h(x,\tau)$ and $\eta(x,\tau)$.
Defining $\delta \equiv \delta \rho /\rho$ for the fluid density perturbations, 
$\theta \equiv \partial_i v^i$ for the divergence of the fluid proper velocity $v^i$ and 
 using Eq.~(\ref{eq:conservDMDE}),  it results, at first order in perturbation theory:
 \begin{eqnarray}
\label{eq:deltambe}
\dot\delta_{dm}  & = & -(\theta_{dm}+\frac12 \dot h)
+\xi {\mathcal H}\frac{\rho_{de}}{\rho_{dm}} (\delta_{de}-\delta_{dm})\, \\
 \label{eq:thetames}
 \dot \theta_{dm}  & = & -{\mathcal H}\theta_{dm} 
+\xi {\mathcal H}(1-b)\frac{\rho_{de}}{\rho_{dm}}(\theta_{de}-\theta_{dm})\,.\\  
\label{eq:deltaees}
\dot\delta_{de}  & = & -(1+w)(\theta_{de}+\frac12 \dot h)
 -3 {\mathcal H}\left(\hat c_{s\,de}^2 -w\right)\left[ \delta_{de} +
{\mathcal H} \left( 3(1+w) + \xi\right)\frac{\theta_{de}}{k^2} \right]\,, \\ \nonumber \\
\label{eq:thetaees}
\dot \theta_{de}  & = & -{\mathcal H}\left(1-3\hat c_{s\,de}^2 -\frac{\hat
  c_{s\,de}^2+b}{1+w} \xi
\right)\theta_{de}+\frac{k^2}{ 1+w}\hat c_{s\,de}^2 \delta_{de}-b\,\xi{\mathcal H}\frac{ \theta_{dm}}{1+w} \,,
\end{eqnarray}
where  $b=1$ stands for models with $Q_\nu\propto u_\nu^{(dm)}$ (\ref{eq:um}) and
$b=0$  for models with $Q_\nu\propto u_\nu^{(de)}$ (\ref{eq:ue}). We
have assumed that ${\mathcal H}$ is the global expansion rate and that
it does not contribute to $Q$ perturbation.

Notice that the continuity equation  for dark matter
(\ref{eq:deltambe}) always includes an extra term compared to
uncoupled models which depends on the dark interaction $Q$ of Eq.~(\ref{eq:Qrhoe}). 
Focusing on viable models, it can be shown\footnote{See
\cite{Koyama:2009gd} for  similar coupled models. The details of the
growth at small scales associated to the models studied here will be
presented elsewhere.}
that the dark matter growth equation at small scales is mainly modified in
two ways: (i) the background evolution (in particular, $\rho_{dm}$ and
$\mathcal H$ evolution) is different, (ii)   the Hubble
friction term and the source term get extra contributions from $Q$.
For negative couplings, these two modifications lead to an enhancement of the dark matter growth compared to
uncoupled models (see also Ref.~\cite{CalderaCabral:2009ja} for similar models). 
The closer $\xi$ gets to -1, the larger is the growth.
This particular feature can be constrained
by cosmological data, in particular in the next section we will see
that large scale structure data provide the strongest limits on the
interaction $Q$.

Also notice that the Euler equation for dark matter~(\ref{eq:thetames}) is only modified
in the $Q_\nu \propto u_\nu^{(de)}$ (\ref{eq:ue}) case, leading a violation of
the weak equivalence principle. Constraints resulting  from
the difference between dark matter and baryon velocities  could provide
additional restrictions on the allowed values of the dark coupling, to be
added to the ones presented in the following section. For the sake of
comparison with the already studied $Q_\nu \propto u_\nu^{(dm)}$
(\ref{eq:um}) model we present below the likelihood plots for the two
coupled models~(\ref{eq:um}) and~(\ref{eq:ue}).

\section{Cosmological constraints from data for $Q=\xi {\mathcal H} \rho_{de}$}
\label{sec:cosmo}
%

%
\begin{figure}[t]
\vspace{-0.1cm}
\begin{tabular}{cc}
\hspace*{-0.75cm} 
\includegraphics[width=8cm]{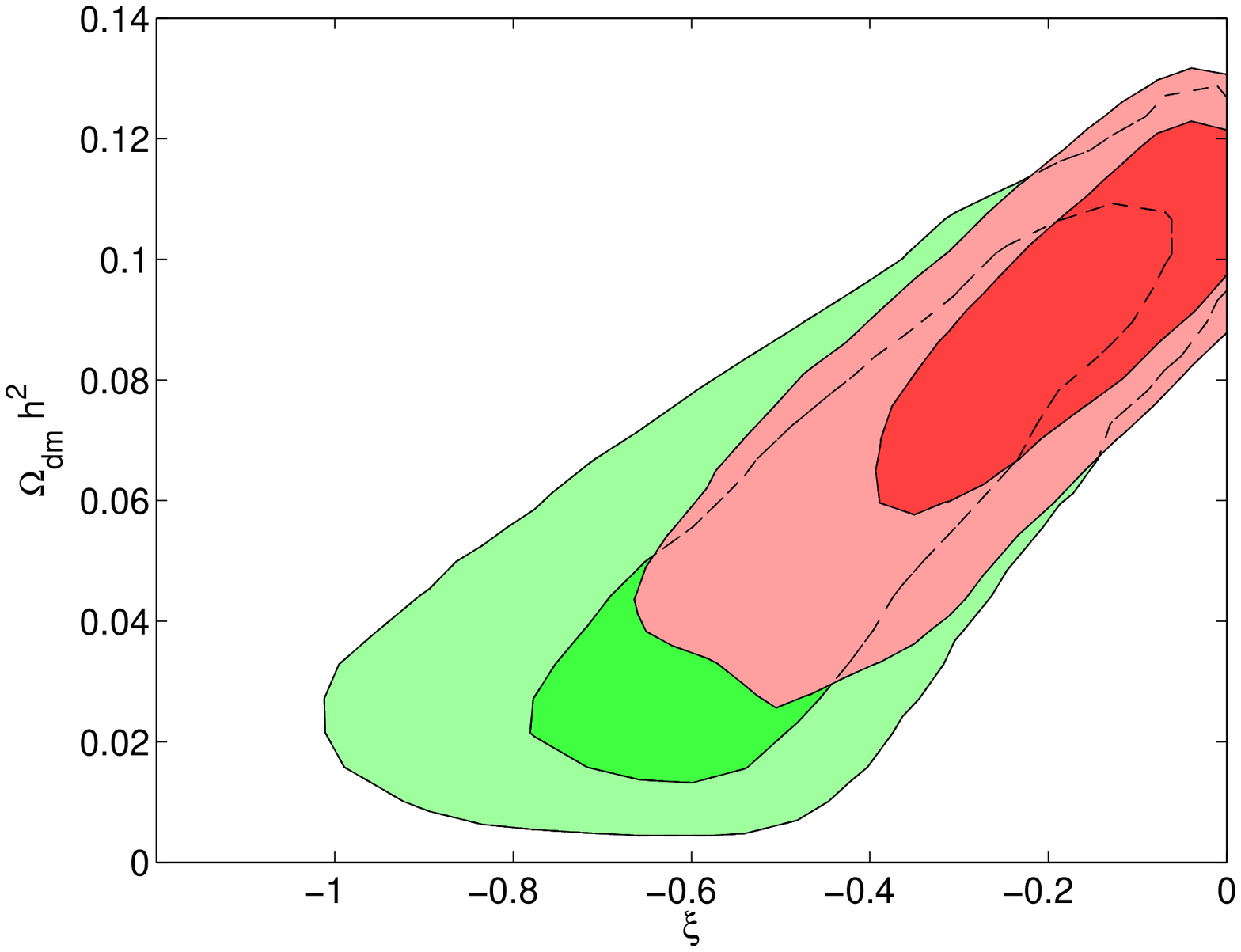} &
\includegraphics[width=8cm]{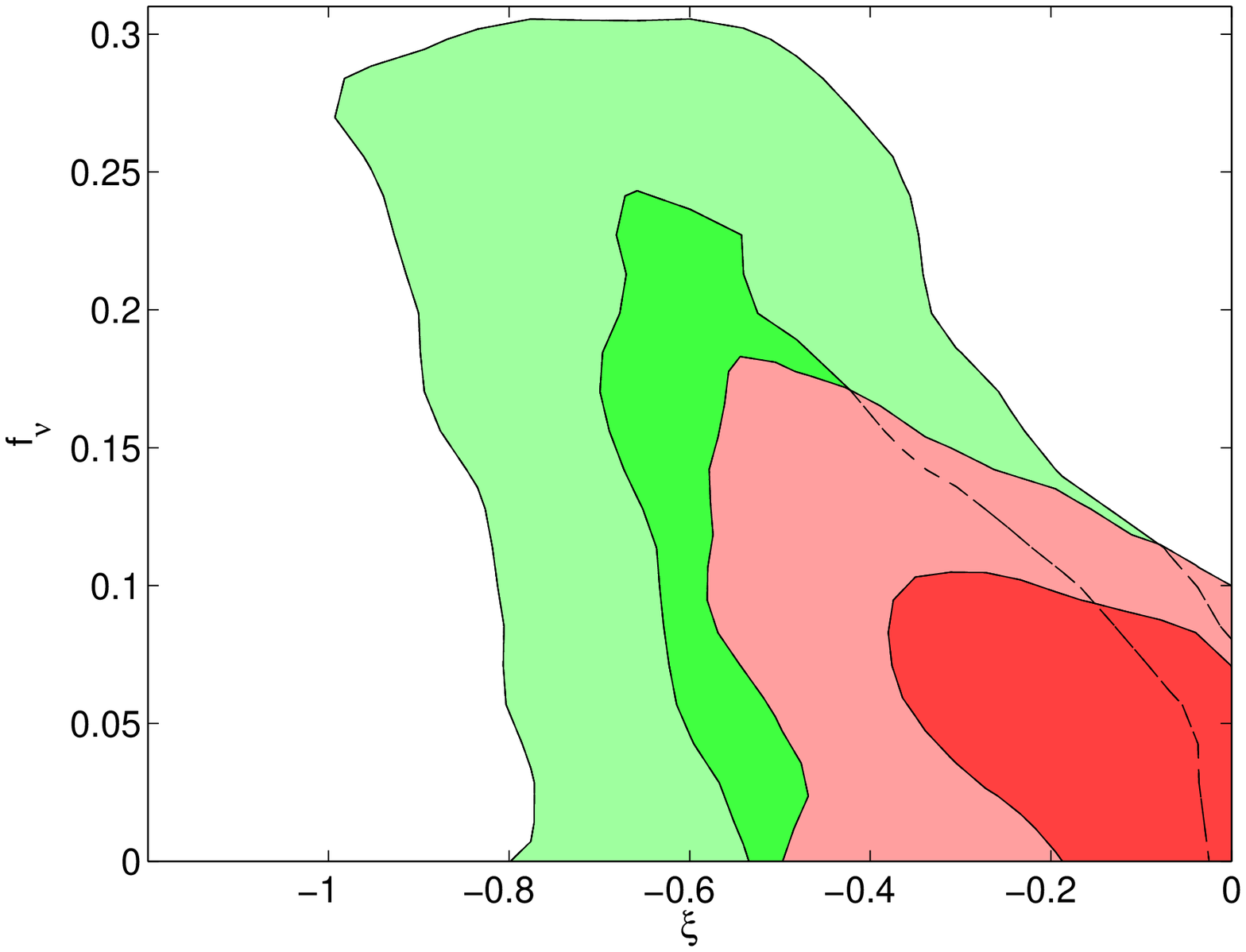} \\
\includegraphics[width=8cm]{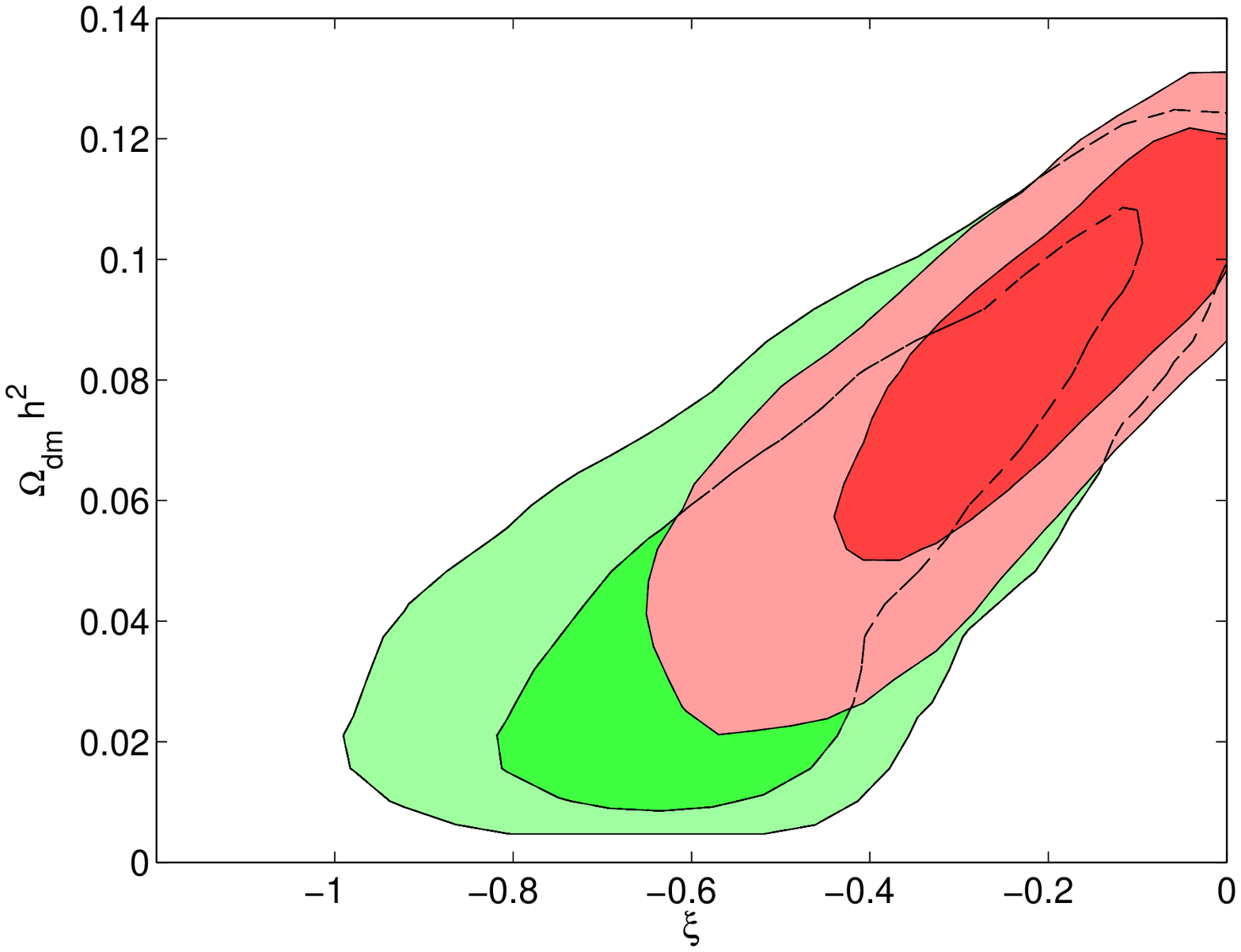} &
\includegraphics[width=8cm]{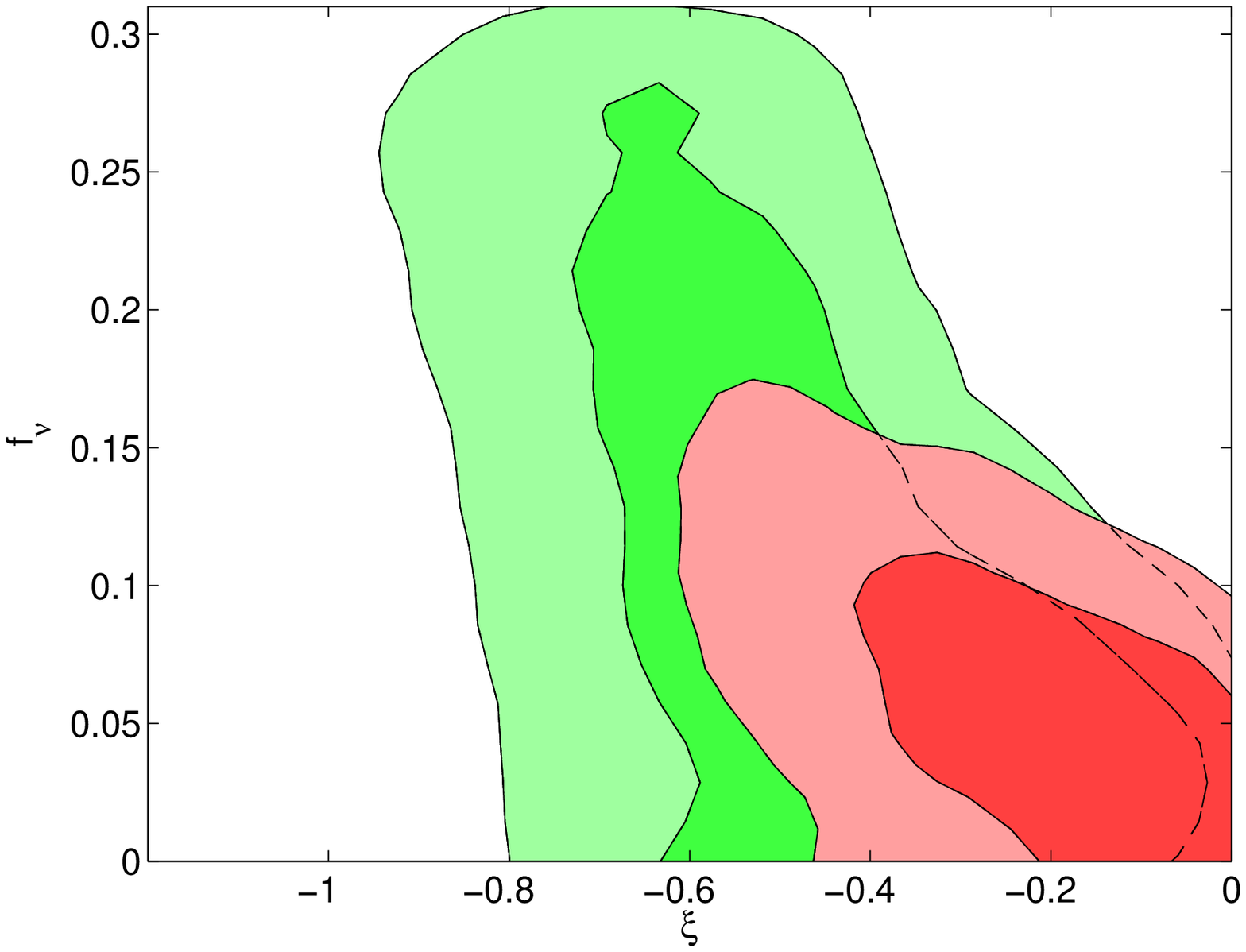} \\
\end{tabular}
\caption{\it Scenario with $Q\propto \rho_{de}$.
The two upper panels correspond to models with $Q_\nu\propto
u_\nu^{(dm)}$, the two lower panels  to models with $Q_\nu\propto
u_\nu^{(de)}$. Left (right) panel: 1$\sigma$ and 2$\sigma$ 
marginalized contours in the $\xi$--$\Omega_{dm} h^2$  ($\xi$--$f_\nu$) plane. The largest, green contours show 
the current constraints from WMAP (5 year data), HST, SN and $H(z)$ data. The smallest, red contours 
show the current constraints from WMAP (5 year data), HST, SN, $H(z)$ and LSS data.}
\label{fig:fig1o}
\end{figure}


In Ref.~\cite{Gavela:2009cy}, we explore  the  constraints on the dark
energy-dark matter coupling $\xi$ using the publicly available package
\texttt{cosmomc}~\cite{Lewis:2002ah}. The latter is  modified in order to
include the coupling among the dark matter and dark energy
components. More details on the cosmological model and
on the priors adopted can be found in Ref.~\cite{Gavela:2009cy}.
The datasets in the analysis are:
\begin{enumerate}
\item WMAP 5-year data~\cite{Dunkley:2008ie,Komatsu:2008hk}
\item  Prior on
the Hubble parameter of $72\pm8$~km/s/Mpc from the Hubble key
project (HST)~\cite{Freedman:2000cf}
\item  Super Novae Ia (SN Ia) data~\cite{Kowalski:2008ez}
\item $H(z)$ data at $0<z<1.8$ from galaxy ages~\cite{Simon:2004tf}
\item Large scale structure data (LSS data)  from the Sloan Digital Sky
Survey~\cite{Tegmark:2006az}
\end{enumerate}
The data analysis is carried out into two runs,  the first run
includes the datasets from 1 to 4 while in the second run the fifth
dataset  is added. We restrict ourselves to $w>-1$ and $\xi <0$, a parameter region which ensures 
a negative doom factor, see Eq.~(\ref{eq:maldito_us}), and thus spans an instability--free 
region of scenarios to explore.  

Figure~\ref{fig:fig1o} (left panel) illustrates the $1$ and $2\sigma$
marginalized contours in the $\xi$--$\Omega_{dm} h^2$ plane, where
$\Omega_{dm}$ is today's ratio between dark matter energy density and
critical energy density.
The results from the two runs described above are shown.
Notice that a huge degeneracy is present, being $\xi$ and $\Omega_{dm}
h^2$ positively correlated.
 The shape of the contours can be easily
understood following our discussion in the previous sections. In
a universe with a negative dark coupling $\xi$, there is an
enhancement of the growth of structure relative to the non interacting
case.  The amount of \emph{intrinsic} dark
matter (which is directly proportional to $ \Omega_{dm} h^2$) needed
to reproduce  the LSS data should decrease as the dark coupling becomes more and
more negative. We also see that the addition of LSS data in the second run (red
contours in Figs.~\ref{fig:fig1o} and~\ref{fig:fig2})  gives the most
stringent constraint on $\xi$.

The right panel of Fig.~\ref{fig:fig1o} shows  the correlation
among the fraction of matter energy-density in the form of massive
neutrinos $f_\nu$ and the dark coupling $\xi$. The relation between
the neutrino fraction used here $f_\nu$ and the neutrino mass for $N_\nu$ degenerate neutrinos reads
\begin{equation}
f_\nu=\frac{\Omega_\nu h^2}{\Omega_{dm} h^2}=\frac{\sum m_\nu}{93.2 \textrm{eV}} \cdot \frac{1}{\Omega_{dm} h^2}=\frac{N_\nu m_\nu}{93.2 \textrm{eV}}\cdot \frac{1}{\Omega_{dm} h^2} ~.
\end{equation}
Neutrinos can indeed play a relevant role in large scale structure
formation and leave key signatures in several cosmological datasets.
Degeneracies between dark energy sector parameters and $\sum m_\nu$ are rather
well known, see Ref.~\cite{Hannestad:2005gj,LaVacca:2008mh,Reid:2009nq}.
Non-relativistic neutrinos in the
recent Universe suppress the growth of matter density fluctuations and
galaxy clustering. This effect can be compensated  by the existence of
a coupling between the dark sectors, due to the fact that in the coupled model
negative couplings enhance the growth of matter density perturbations.

Notice that the constraints on the $Q_\nu\propto u_\nu^{(dm)}$ (\ref{eq:um}) and
 $Q_\nu\propto u_\nu^{(de)}$ (\ref{eq:ue}) models are rather
equivalent, due to the fact that the background evolution history (see Eqs.~(\ref{eq:adm}) and~(\ref{eq:ade})) is the same, and the evolution of perturbations~(\ref{eq:deltambe})~-~(\ref{eq:thetaees}) is also very similar for these two models.


\section{Conclusions}
\label{sec:concl}
%
\begin{figure}[t]
\vspace{-0.1cm}
\begin{tabular}{cc}
\includegraphics[width=8cm]{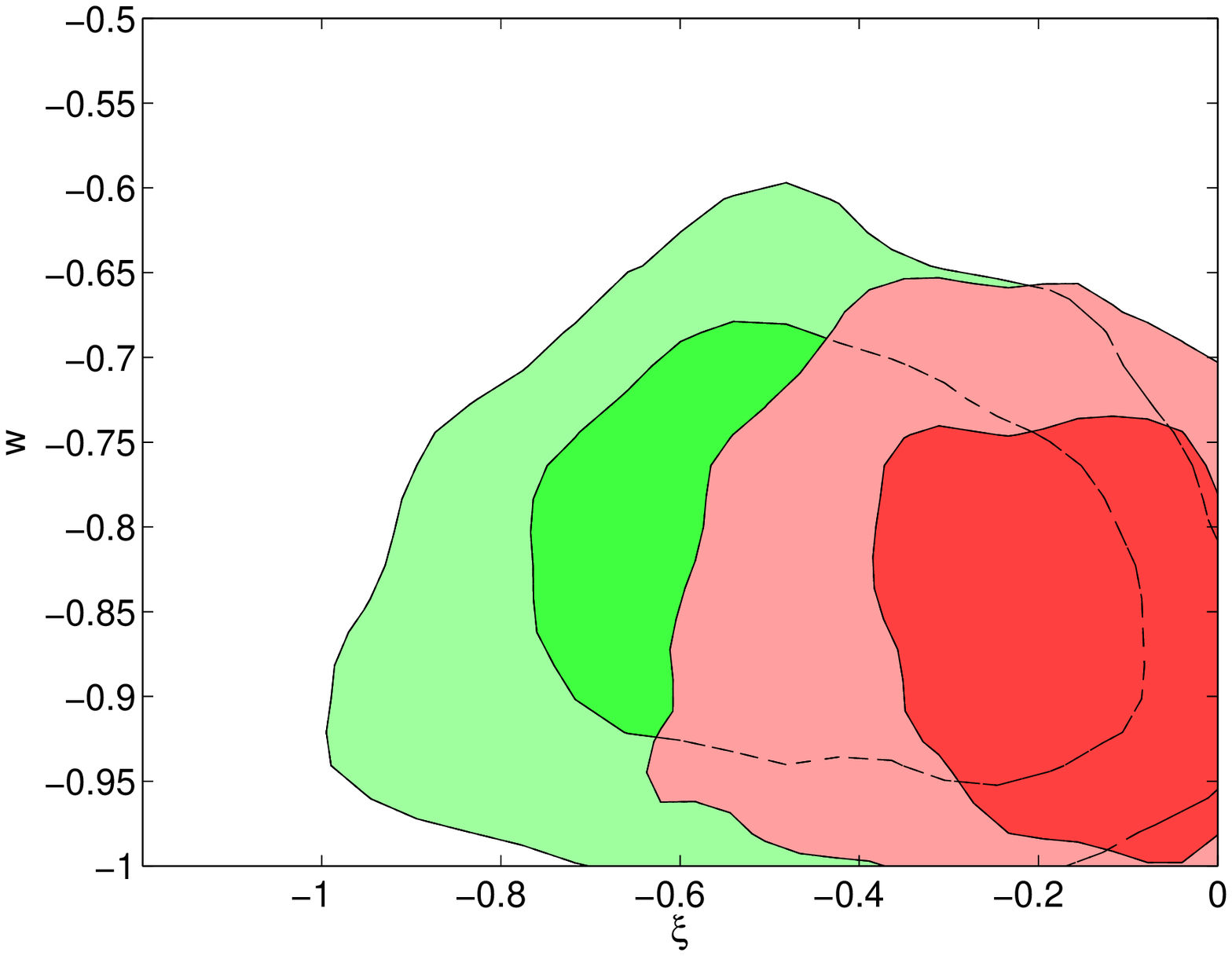} &
\includegraphics[width=8cm]{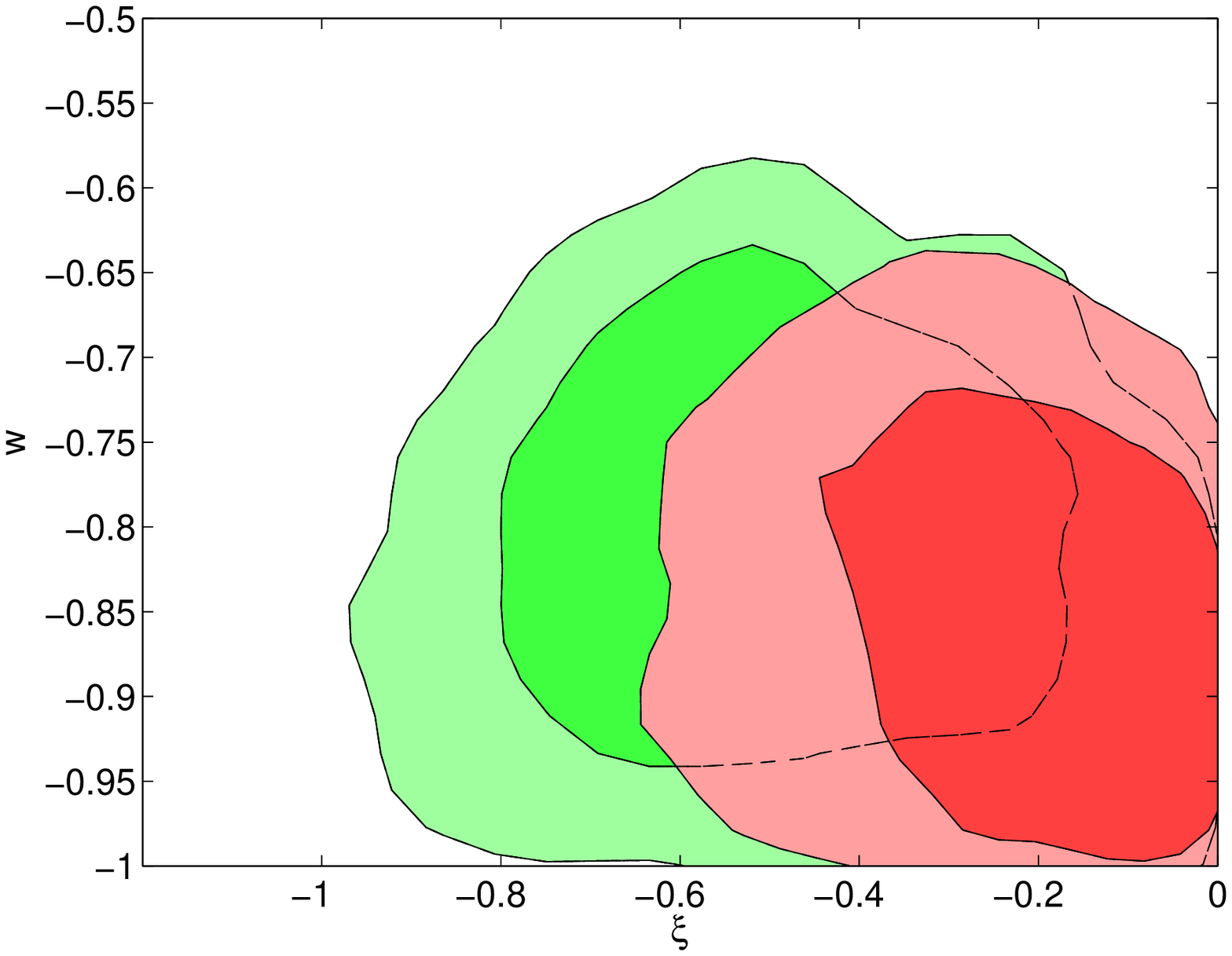} \\
\end{tabular}
\caption{\it Scenario with $Q\propto \rho_{de}$. Left (right) panel: 1$\sigma$ and 2$\sigma$ 
marginalized contours in the $\xi$--$w$  plane for $Q_\nu\propto
u_\nu^{(dm)}$  ($Q_\nu\propto
u_\nu^{(dm)}$). The largest, green contours show 
the current constraints from WMAP (5 year data), HST, SN and $H(z)$ data. The smallest, red contours 
show the current constraints from WMAP (5 year data), HST, SN, $H(z)$ and LSS data.}
\label{fig:fig2}
\end{figure}


In these proceedings, we discuss the origin of non-adiabatic instabilities in
coupled models satisfying \\
$\nabla_\mu T^\mu_{dm\,\nu}= Q u_\nu^{dm\,(de)}/a $ and  $\nabla_\mu T^\mu_{de\,\nu}= Q u_\nu^{dm\,(de)}/a $.
We show that the
sign of the doom factor 
\begin{equation}
{\bf d} \equiv \frac{Q}{3\mathcal H\rho_{de}(1+w)}
\end{equation}
reveals the presence
 of instabilities. 
In particular, when ${\bf d }$ is positive and sizeable, ${\bf d }>1$, the dark-coupling 
dependent terms may dominate the evolution of the  dark energy
perturbation, and drive an  unstable growth regime. Notice that {\bf d} has been defined for a constant dark energy
 equation of state $w$ (see {\it e.g.} Ref.~\cite{Majerotto:2009np,
   Valiviita:2009nu} for $w=w(a)$). 

 We have  studied the constraints from current cosmological data on the dimensionless coupling $\xi$ in a class of viable models in which $Q=\xi {\mathcal H} \rho_{de}$, independently of the dark interaction term 4-velocity dependence. The analyses presented here were carried out in the  $\xi<0$ and positive 
$(1+w)$ region of the parameter space, which offers the best agreement with data on large scale 
structure formation. From the results of our fits we find that both $w$ and $\xi$ are not very
constrained from data: it can be noticed from Fig.~\ref{fig:fig2} that substantial values 
for both parameters, near -0.5, are easily allowed. Furthermore, 
$\xi$ turns out to be positively correlated with  $\Omega_{dm}
h^2$ and a larger neutrino fraction $f_{\nu}$ than in uncoupled models is allowed for negative values of the coupling $\xi$.


\begin{theacknowledgments}
The work reported associated to the  $Q_\nu\propto u_\nu^{(dm)}$ models has been done in collaboration with B.~Gavela,
D.~Hernandez, O.~Mena and S.~Rigolin, see Ref.~\cite{Gavela:2009cy}. L.~L.~H
thanks the organizers of the Invisible Universe conference (Paris) for
giving her the opportunity to give a talk on ``dark couplings'' and the I.F.I.C. group
(Valencia), where part of the $Q_\nu\propto u_\nu^{(de)}$ work  
was carried out, for its hospitality .  We also acknowledge
R. Jim\'enez B.~Reid and L. Verde for useful comments and discussions.
 L.~L.~H was partially supported by CICYT through
the project FPA2006-05423, by CAM through the project HEPHACOS, P-ESP-00346,  by the PAU (Physics
of the accelerating universe) Consolider Ingenio 2010, by the
F.N.R.S. and  the I.I.S.N.. O.~M. work is supported by a Ram\'on y Cajal contract from MEC, Spain

\end{theacknowledgments}

\appendix
\section{Appendix: Growth of perturbations in strongly coupled scenarios}
\label{a:grstrongfull}
 The strong coupling regime can be characterized by 
 \begin{eqnarray}
\left|\frac{Q}{\mathcal H\rho_{de}}\right|&\gg&\left|3(1+w)\right| \,,
  \label{eq:condstr_full}\\
\left |\frac{Q}{\mathcal H\rho_{de}}\,\frac{\hat c_{s\,de}^2+1}{1+w }\right|  &\gg& \left|1-3\hat c_{s\,de}^2\right|\,,
\end{eqnarray}
which ensure that the dark-coupling terms dominate the evolution of both $\delta_{de}$ and $\theta_{de}$. 
With $c_{s\,de}^2>0$, Eq.~(\ref{eq:condstr_full}) alone is enough to define the regime.

For coupled models satisfying Eqs.~(\ref{eq:conservDMDE}),~(\ref{eq:um})
and~(\ref{eq:ue}), the resulting growth equation for dark energy perturbations at large
scales ($\mathcal H/k \gg 1$) can be approximated by
\begin{eqnarray}
  \label{eq:grstrongfullc}
  \delta_{de}''&\simeq&  \frac{\delta_{de}'}{a}\left(\frac Q{{\mathcal H}\rho_{de}} \,
  \frac{\hat c_{s\,de}^2+b}{1+w}+ a \left(\ln[Q/\rho_{de}]
  \right)'\right)\cr
&&+3\frac{\delta_{de}}{a^2}(\hat c_{s\,de}^2-w)
\left(\frac
    Q{{\mathcal H}\rho_{de}} \,
  \frac{b}{1+w}+ a \left(\ln[Q/\rho_{de}] \right)'\right)\cr
&&+\frac {1}{a^2{\mathcal H}}\delta[Q/\rho_{de}]
\left(\frac{\hat c_{s\,de}^2+b}{1+w}\frac Q{{\mathcal
      H}\rho_{de}}+a \left(\ln[Q/\rho_{de}] \right)'-\frac12-\frac32 w
\Omega_{de}\right)\cr
&&-\frac {1}{a {\mathcal H}}\left(\delta[Q/\rho_{de}]\right)' -(1+w)\frac{\dot h}2\,.
\end{eqnarray}
where $b=1$ for models with $Q_\nu\propto u_\nu^{(dm)}$ (\ref{eq:um}) and
$b=0$  for models with $Q_\nu\propto u_\nu^{(de)}$ (\ref{eq:ue}).
  
%



\bibliographystyle{aipproc}   
\bibliography{bibdmde-v2.bib}

\begin{thebibliography}{28}
\expandafter\ifx\csname natexlab\endcsname\relax\def\natexlab#1{#1}\fi
\providecommand{\enquote}[1]{``#1''}
\expandafter\ifx\csname url\endcsname\relax
  \def\url#1{\texttt{#1}}\fi
\expandafter\ifx\csname urlprefix\endcsname\relax\def\urlprefix{URL }\fi
\providecommand{\eprint}[2][]{\url{#2}}

\bibitem[Caldera-Cabral et~al.(2009)]{CalderaCabral:2009ja}
G.~Caldera-Cabral, R.~Maartens, and B.~M. Schaefer, \emph{JCAP} \textbf{0907},
  027 (2009), \eprint{0905.0492}.

\bibitem[Valiviita et~al.(2008)]{Valiviita:2008iv}
J.~Valiviita, E.~Majerotto, and R.~Maartens, \emph{JCAP} \textbf{0807}, 020
  (2008), \eprint{0804.0232}.

\bibitem[Koyama et~al.(2009)]{Koyama:2009gd}
K.~Koyama, R.~Maartens, and Y.-S. Song  (2009), \eprint{0907.2126}.

\bibitem[Wetterich(1995)]{Wetterich:1994bg}
C.~Wetterich, \emph{Astron. Astrophys.} \textbf{301}, 321--328 (1995),
  \eprint{hep-th/9408025}.

\bibitem[Amendola(1999)]{Amendola:1999qq}
L.~Amendola, \emph{Phys. Rev.} \textbf{D60}, 043501 (1999),
  \eprint{astro-ph/9904120}.

\bibitem[Jackson et~al.(2009)]{Jackson:2009mz}
B.~M. Jackson, A.~Taylor, and A.~Berera, \emph{Phys. Rev.} \textbf{D79}, 043526
  (2009), \eprint{0901.3272}.

\bibitem[He et~al.(2009)]{He:2008si}
J.-H. He, B.~Wang, and E.~Abdalla, \emph{Phys. Lett.} \textbf{B671}, 139--145
  (2009), \eprint{0807.3471}.

\bibitem[Gavela et~al.(2009)]{Gavela:2009cy}
M.~B. Gavela, D.~Hernandez, L.~L. Honorez, O.~Mena, and S.~Rigolin, \emph{JCAP}
  \textbf{0907}, 034 (2009), \eprint{0901.1611}.

\bibitem[Majerotto et~al.(2009)]{Majerotto:2009np}
E.~Majerotto, J.~Valiviita, and R.~Maartens  (2009), \eprint{0907.4981}.

\bibitem[Corasaniti(2008)]{Corasaniti:2008kx}
P.~S. Corasaniti, \emph{Phys. Rev.} \textbf{D78}, 083538 (2008),
  \eprint{0808.1646}.

\bibitem[Chongchitnan(2009)]{Chongchitnan:2008ry}
S.~Chongchitnan, \emph{Phys. Rev.} \textbf{D79}, 043522 (2009),
  \eprint{0810.5411}.

\bibitem[Dunkley et~al.(2009)]{Dunkley:2008ie}
J.~Dunkley, et~al., \emph{Astrophys. J. Suppl.} \textbf{180}, 306--329 (2009),
  \eprint{0803.0586}.

\bibitem[Komatsu et~al.(2009)]{Komatsu:2008hk}
E.~Komatsu, et~al., \emph{Astrophys. J. Suppl.} \textbf{180}, 330--376 (2009),
  \eprint{0803.0547}.

\bibitem[Freedman et~al.(2001)]{Freedman:2000cf}
W.~L. Freedman, et~al., \emph{Astrophys. J.} \textbf{553}, 47--72 (2001),
  \eprint{astro-ph/0012376}.

\bibitem[Kowalski et~al.(2008)]{Kowalski:2008ez}
M.~Kowalski, et~al., \emph{Astrophys. J.} \textbf{686}, 749--778 (2008),
  \eprint{0804.4142}.

\bibitem[Simon et~al.(2005)]{Simon:2004tf}
J.~Simon, L.~Verde, and R.~Jimenez, \emph{Phys. Rev.} \textbf{D71}, 123001
  (2005), \eprint{astro-ph/0412269}.

\bibitem[Tegmark et~al.(2006)]{Tegmark:2006az}
M.~Tegmark, et~al., \emph{Phys. Rev.} \textbf{D74}, 123507 (2006),
  \eprint{astro-ph/0608632}.

\bibitem[Lewis et~al.(2000)]{Lewis:1999bs}
A.~Lewis, A.~Challinor, and A.~Lasenby, \emph{Astrophys. J.} \textbf{538},
  473--476 (2000), \eprint{astro-ph/9911177}.

\bibitem[Lewis and Bridle(2002)]{Lewis:2002ah}
A.~Lewis, and S.~Bridle, \emph{Phys. Rev.} \textbf{D66}, 103511 (2002),
  \eprint{astro-ph/0205436}.

\bibitem[Kodama and Sasaki(1984)]{Kodama:1985bj}
H.~Kodama, and M.~Sasaki, \emph{Prog. Theor. Phys. Suppl.} \textbf{78}, 1--166
  (1984).

\bibitem[Afshordi et~al.(2005)]{Afshordi:2005ym}
N.~Afshordi, M.~Zaldarriaga, and K.~Kohri, \emph{Phys. Rev.} \textbf{D72},
  065024 (2005), \eprint{astro-ph/0506663}.

\bibitem[Kaplinghat and Rajaraman(2007)]{Kaplinghat:2006jk}
M.~Kaplinghat, and A.~Rajaraman, \emph{Phys. Rev.} \textbf{D75}, 103504 (2007),
  \eprint{astro-ph/0601517}.

\bibitem[Bean et~al.(2008)]{Bean:2007ny}
R.~Bean, E.~E. Flanagan, and M.~Trodden, \emph{Phys. Rev.} \textbf{D78}, 023009
  (2008), \eprint{0709.1128}.

\bibitem[Ma and Bertschinger(1995)]{Ma:1995ey}
C.-P. Ma, and E.~Bertschinger, \emph{Astrophys. J.} \textbf{455}, 7--25 (1995),
  \eprint{astro-ph/9506072}.

\bibitem[Hannestad(2005)]{Hannestad:2005gj}
S.~Hannestad, \emph{Phys. Rev. Lett.} \textbf{95}, 221301 (2005),
  \eprint{astro-ph/0505551}.

\bibitem[La~Vacca et~al.(2009)]{LaVacca:2008mh}
G.~La~Vacca, S.~A. Bonometto, and L.~P.~L. Colombo, \emph{New Astron.}
  \textbf{14}, 435--442 (2009), \eprint{0810.0127}.

\bibitem[Reid et~al.(2009)]{Reid:2009nq}
B.~A. Reid, L.~Verde, R.~Jimenez, and O.~Mena  (2009), \eprint{0910.0008}.

\bibitem[Valiviita et~al.(2009)]{Valiviita:2009nu}
J.~Valiviita, R.~Maartens, and E.~Majerotto  (2009), \eprint{0907.4987}.

\end{thebibliography}

\end{document}